\documentclass[12pt,english,floatfix,showkeys,superscriptaddress,aps,prd,preprint]{revtex4}
\usepackage[latin1]{inputenc}
\usepackage[T1]{fontenc}
\usepackage{lmodern}
\usepackage{amsmath}
\usepackage{amssymb}
\usepackage{graphicx}
\usepackage{float}
\usepackage{esint}
\usepackage{dcolumn}
\usepackage{babel}
\usepackage{csquotes}
\usepackage{color}
\usepackage{caption}
\usepackage{subcaption}
\usepackage{slashed}
\usepackage{simplewick}
\usepackage{tikz-feynman}
\usepackage[]{tikz-feynman}
\usepackage{amsmath,latexsym}

\usepackage{cancel}

\usepackage{hyperref}
\hypersetup{
    colorlinks,
    citecolor=blue,
    filecolor=green,
    linkcolor=purple,
    urlcolor=red,
}

\usepackage{slashed}

\usepackage{hyperref}
\hypersetup{colorlinks,breaklinks,
			citecolor=[rgb]{0,0.0,1.0},
            urlcolor=[rgb]{0.0,0.0,1.0},
            linkcolor=[rgb]{0,0.5,0.9}}

\begin{document}

\title{Meson scattering in a non-minimally Lorentz-violating scalar QED at finite temperature}

\author{M. C. Ara\'{u}jo}
\email{michelangelo@fisica.ufc.br}
\affiliation{Universidade Federal do Cear\'a (UFC), Departamento de F\'isica,\\ Campus do Pici, Fortaleza - CE, C.P. 6030, 60455-760 - Brazil.}
\author{R. V. Maluf}
\email{r.v.maluf@fisica.ufc.br}
\affiliation{Universidade Federal do Cear\'a (UFC), Departamento de F\'isica,\\ Campus do Pici, Fortaleza - CE, C.P. 6030, 60455-760 - Brazil.}
\author{J. Furtado}
\email{job.furtado@ufca.edu.br}
\affiliation{Universidade Federal do Cariri (UFCA), Av. Tenente Raimundo Rocha, \\ Cidade Universit\'{a}ria, Juazeiro do Norte, Cear\'{a}, CEP 63048-080, Brasil}
\affiliation{Department of Physics, Faculty of Science, Gazi University, 06500 Ankara, Turkey}



\date{\today}

\begin{abstract}
In this paper we study meson scattering in a non-minimally Lorentz-violating scalar QED at finite temperature. The meson scatterings were investigated in tree level and the finite temperature effects were addressed by using the thermofield dynamics formalism. We have considered three types of scattering, namely, meson-antimeson of $a$-type into meson-antimeson of $b$-type, meson-antimeson of $a$-type into meson-antimeson of $a$-type and meson-meson of $a$-type into meson-meson of $a$-type. For each scattering we have computed the cross section in order to investigate the influence of the finite temperature effects.
\end{abstract}


\maketitle

\section{Introduction}\label{intro}

In recent years, there has been a focus on exploring potential expansions of the Standard Model (SM). Within this framework, the discussion has shifted towards the breaking of Lorentz and CPT symmetries, considered as a significant subject \cite{Kos1, Kos2, Kos3, Kos4, Colladay:1996iz, Colladay:1998fq}. Typically, Lorentz symmetry is disrupted by introducing specific directions in space-time, represented by additive terms that are proportionate to small constant vectors or tensors. The Standard Model Extension (SME) \cite{Colladay:1996iz, Colladay:1998fq} stands out as the most established model addressing the consequences of Lorentz and CPT symmetry violation. SME, being an effective field theory, incorporates all conceivable terms violating Lorentz and CPT symmetries in its Lagrangian.

Kostelecky \cite{Edwards:2018lsn}. proposed an extension of the scalar sector incorporating Lorentz-violating effects. This model presents a comprehensive effective scalar field theory across any spacetime dimension, featuring explicit perturbative spin-independent Lorentz-violating operators of arbitrary mass dimension. The significance of this development lies in the predominance of spin in most fundamental particles of the SM, with the Higgs boson being the sole example of a fundamental spinless particle. Despite the relatively minor role of the scalar sector of QED (sQED) compared to strong interactions in describing meson coupling, it has been suggested \cite{Edwards:2019lfb} that a Lorentz-violating extension of sQED could effectively address slight CPT deviations in neutral-meson oscillations.

The minimal extension of SME was extensively studied in the last years in several contexts, such as radiative corrections \cite{Jackiw:1999yp, Kostelecky:2001jc, Ferrari:2021eam}, neutrinos oscillations \cite{MINOS:2008fnv, MINOS:2010kat, MiniBooNE:2011pix}, Euler-Heisenberg effective action \cite{Ferrari:2021eam, Furtado:2014cja}, gravitational context \cite{Kostelecky:2016kkn, Assuncao:2018jkq}, finite temperature field theory \cite{Assuncao:2018jkq, Leite:2013pca, Leite:2011jg, Assuncao:2016fko, Araujo:2023izx} among others (for a more complete review, see f.e. \cite{Mariz:2022oib}). The non-minimal extension, despite its posses only nonrenormalizable terms, have been receiving attention in the literature \cite{Myers:2003fd, Kostelecky:2009zp, Kostelecky:2011gq, Mariz:2010fm, Rubtsov:2012kb}. A strong motivation lies in the fact that some relevant astrophysical processes impose severe restrictions on the coefficients associated with the operators with dimension $d \geq 5$, being such contributions comparable or even dominant when compared with the ones that arise from dimension $d \leq 4$ operators. Within the nonminimal LV framework, an important role was played by the paper \cite{Kostelecky:2009zp}, in which the simplest cases of such operators were, for the very first time, introduced in the LV scenario. The first studies of their perturbative impact were investigated in \cite{Gomes:2009ch, Mariz:2010fm, Gazzola:2010vr, Mariz:2011ed, BaetaScarpelli:2013rmt}. And finally, all possible LV extensions of SME with dimensions up to 6 for fermion-dependent operators were listed in \cite{Kostelecky:2018yfa}.

The topological structure of the thermofield dynamics (TFD) formalism allows the study of thermal effects in several different systems, ranging from particle physics to quantum computing and black hole physics \cite{Araujo:2022qke, Ulhoa:2023opw, quantumcomp1}. The central points of the TFD formalism are the duplication of the Hilbert space and the use of Bogoliubov transformation \cite{Umezawa1, Umezawa2}. The thermal Hilbert space (or doubled Hilbert space) allows the construction of a so called thermal ground state. Such ingredients are related to an important result in which it is shown that for any arbitrary operator, its statistical avarage is equal to its vacuum expectation value. In the context of Lorentz symmetry violation, the TFD formalism is being widely and successfully used for computing the finite temperature effects of tree level (while finite temperature loop corrections are usully computed using the Matsubara formalism \cite{Araujo:2023izx, Assuncao:2020cdo}) scatterings for QED processes in both spinorial \cite{Cabral:2024tqa, Santos:2021hed, Souza:2021uuf} and scalar sectors \cite{Araujo:2022qke}. Moreover, the thermal effects of Lorentz-violating extension of non-Abelian sectors were also investigated in the TFD formalism \cite{Santos:2022zfz}.  

In this paper we study meson scattering in a non-minimally Lorentz-violating scalar QED at finite temperature. The meson scatterings were investigated in tree level and the finite temperature effects were addressed by using the thermofield dynamics formalism. We have considered three types of scattering, namely, meson-antimeson of $a$-type into meson-antimeson of $b$-type, meson-antimeson of $a$-type into meson-antimeson of $a$-type and meson-meson of $a$-type into meson-meson of $a$-type. For each scattering we have computed the cross section in order to investigate the influence of the finite temperature effects.
    
This paper is organized as follows. In Sec. \ref{secdois}, a revision of the thermofield dynamics formalism is presented. In Sec. \ref{sec3}, we discuss the non-minimally Lorentz-violating scalar sector of QED we are considering in both zero and finite temperature. In Sec. \ref{sec4} we calculate the differential cross sections for mesons at finite temperature by employing the TFD formalism. Finally, in Sec. \ref{conclusion}, the main results of the paper are summarized. Throughout the paper, we have used the natural units $\hslash = c = k_B = 1$, with $k_B$ being the Boltzmann constant, and assumed $\eta^{\mu\nu}=diag(1,-1,-1,-1)$ as the Minkowski metric.

\section{Thermofield Dynamics Formalism}\label{secdois}

The Thermofield Dynamics (TFD) is a real-time formalism that allows the investigation of temperature effects on systems described by a quantum field theory. Its main feature is the definition of a thermal vacuum state $|\, 0(\beta)\, \rangle$ through the duplication of the degrees of freedom of the system, characterized by the double-space $\hat{\mathcal{S}} = \mathcal{S}\otimes \Tilde{\mathcal{S}}$. Here, $\Tilde{S}$ is to be thought of as a copy of the original space $S$, where all the physical observables are actually defined, and $\beta = T^{-1}$, being $T$ the system temperature. Due to this doubled structure, for each operator $\mathcal{O}$ in the space $\mathcal{S}$, there exists an operator $\hat{\mathcal{O}}$ in $\hat{\mathcal{S}}$ defined by
\begin{eqnarray}
\hat{\mathcal{O}} \equiv \mathcal{O} - \Tilde{\mathcal{O}},
\end{eqnarray} where $\Tilde{\mathcal{O}}$ is the analogous operator in the tilde-space $\Tilde{\mathcal{S}}$. Tilde and non-tilde operators are connected by the following conjugation rules:
\begin{eqnarray}\label{rct1}
\widetilde{(\mathcal{O}_i\mathcal{O}_j)} &=& \tilde{\mathcal{O}_i}\tilde{\mathcal{O}_j} ,\\
\widetilde{(a\mathcal{O}_i+b\mathcal{O}_j)} &=& a^{\ast}\Tilde{\mathcal{O}_i}+b^{\ast}\Tilde{\mathcal{O}_j},\\
(\Tilde{\mathcal{O}_i})^{\dagger} &=& \widetilde{(\mathcal{O}_i^{\dagger})},\\
\label{rct4}
\widetilde{(\Tilde{\mathcal{O}}_i)} &=& \kappa \mathcal{O}_i,
\end{eqnarray} where $\kappa=1\, (-1)$ for bosons (fermions). As a consequence, thermal operators can be defined through Bogoliubov transformations, which consist of a rotation in the tilde and non-tilde variables. For bosons, for example, the relations between thermal and non-thermal operators are given by 
\begin{eqnarray}
\label{oprtermoperzero1}
a(\mathbf{k},\beta) &=& u_B(\mathbf{k},\beta)a(\mathbf{k}) - v_B(\mathbf{k},\beta)\tilde{a}^{\dagger}(\mathbf{k}) ,\\ \label{oprtermoperzero2}
a^{\dagger}(\mathbf{k},\beta) &=& u_B(\mathbf{k},\beta)a^{\dagger}(\mathbf{k}) - v_B(\mathbf{k},\beta)\tilde{a}(\mathbf{k}), \\ \label{oprtermoperzero3}
\tilde{a}(\mathbf{k},\beta) &=& u_B(\mathbf{k},\beta)\tilde{a}(\mathbf{k}) - v_B(\mathbf{k},\beta)a^{\dagger}(\mathbf{k}),\\ \label{oprtermoperzero4}
\tilde{a}^{\dagger}(\mathbf{k},\beta) &=& u_B(\mathbf{k},\beta)\tilde{a}^{\dagger}(\mathbf{k}) - v_B(\mathbf{k},\beta)a(\mathbf{k}),
\end{eqnarray} where $a\, (\Tilde{a})$ and $a^{\dagger}\, (\Tilde{a}^{\dagger})$ stand for an annihilation and creation operators in the space $\mathcal{S}$ ($\Tilde{\mathcal{S}}$), respectivally. Above, 
\begin{eqnarray}\label{funcaoubossoncosh1}
u_{B}(\mathbf{k},\beta) &=& e^{\beta \, |E(\mathbf{k})|/2}v_B(\mathbf{k},\beta),
\end{eqnarray} and
\begin{eqnarray}
    \label{funcaoubossoncosh2}
v_{B}(\mathbf{k},\beta) &=& (e^{\beta\,  |E(\mathbf{k})|}-1)^{-1/2},
\end{eqnarray} are the functions that carry the temperature dependence of the thermal operators. Here, $E(\mathbf{k})$  is the energy of oscillation associated with the mode $\mathbf{k}$. Also, it can be shown that thermal operators must satisfy
\begin{eqnarray}
    \left[ a(\mathbf{k},\beta),a^{\dagger}(\mathbf{k'},\beta)\right] &=& (2\pi)^3\delta^3(\mathbf{k}-\mathbf{k'}),
\end{eqnarray} with the same conventional algebra for tilde operators and all the other commutation relations being zero. 

In the TFD formalism, when written in the center-of-mass reference frame, the differential cross section takes the form of
\begin{eqnarray}\label{secdifbetacmsomaspinmhatcmbetaquadrado}
      \left( \frac{d \sigma }{d\Omega}   \right)_{\beta,cm} = \frac{1}{2E^2_{cm}}\frac{ | \mathbf{k} |}{16 \pi^2 E_{cm}}|\hat{\mathcal{M}}(\beta)|^2,
\end{eqnarray} where
\begin{equation}\label{matrizshatbeta}
   \hat{\mathcal{M}}(\beta) \equiv \sum_{n=0}^{\infty}\frac{(-i)^n}{n!}\int\, d^4z_1\cdots d^4z_n\, {}_{\beta}\langle \, f\, |\mathcal{T}\left[ \hat{\mathcal{L}}_{int}(z_1)\cdots \hat{\mathcal{L}}_{int}(z_n) \right]|\, i\, \rangle_{\beta},
\end{equation} defines the temperature-dependent transition amplitude for a scattering of interest. In expressions \eqref{secdifbetacmsomaspinmhatcmbetaquadrado} and \eqref{matrizshatbeta}, $E_{cm}$ is the center-of-mass energy, $\mathbf{k}$ is the $3$-momentum of an outgoing particle, $\hat{\mathcal{L}}_{int}(z)$ is the interaction Lagrangian density in the space $\hat{\mathcal{S}}$, and $\mathcal{T}$ is the time-ordering operator. From now on, the procedure to be taken is quite standard in quantum field theory, however, we must make use of the Bogoliubov transformations \eqref{oprtermoperzero1}-\eqref{oprtermoperzero4} in order to first write Eq. \eqref{matrizshatbeta} in terms of the thermal creation and annihilation operators for some Hamiltonian of interest. This will allow us to deal properly with the initial and final states of particles respectively defined by 
\begin{eqnarray}
  |\, i\, \rangle_{\beta} &\equiv& \sqrt{2E(\mathbf{k}_1) 2E(\mathbf{k}_2)}\, a^{\dagger}(\mathbf{k}_1,\beta)a^{\dagger}(\mathbf{k}_2,\beta)\, |\, 0(\beta)\, \rangle,
\end{eqnarray} and 
\begin{eqnarray}
    {}_{\beta}\langle \, f\, | &\equiv& \langle \, 0(\beta)\, | \, a(\mathbf{k}_3,\beta)a(\mathbf{k}_4,\beta) \sqrt{2E(\mathbf{k}_3) 2E(\mathbf{k}_4)},
\end{eqnarray} where $E(\mathbf{k}_i) = \sqrt{\mathbf{k}_i^2+m^2}$ is the energy associated with the mode $\mathbf{k}_i$. 

\section{Non-minimally Lorentz-violating scalar QED}\label{sec3}

The model we are interested in, which will allow us to investigate possible effects of Lorentz symmetry violation in meson scattering, is given by the Lagrangian density
\begin{eqnarray}\label{densmodelgeneral}
\mathcal{L} =(D_{\mu}\phi)^{*}(D^{\mu}\phi)-m^2\phi^{*}\phi-\frac{1}{4}F^{\mu\nu}F_{\mu\nu},
\end{eqnarray} where 
\begin{eqnarray}\label{derivcovarinpnminimal}
D_{\mu} =  \partial_{\mu}-ieA_{\mu}-i\frac{g}{2}\epsilon_{\mu\nu\alpha\beta}\omega^{\nu}F^{\alpha\beta}
\end{eqnarray} is a covariant derivative allowing for non-minimal coupling between the complex scalar field and the gauge field. Supposed to be constant, the vector $\omega^{\mu}$ sets a preferential direction in spacetime and thus breaks the equivalence between particle and observer transformations, thereby promoting the violation of Lorentz symmetry. By substituting Eq. \eqref{derivcovarinpnminimal} into \eqref{densmodelgeneral}, we can conclude that the only modifications to the usual Feynman's rules of the scalar QED come from the terms that make up the Lagrangian density of interaction $\mathcal{L}_{int}$:
\begin{eqnarray}\label{lagintdoistermos}
 \mathcal{L}_{int} = ieA_{\mu}(\phi^{\ast}\partial^{\mu}\phi - \partial^{\mu}\phi^{\ast}\phi  ) + ig \epsilon^{\mu \nu \alpha \beta} \omega_{\nu}\partial_{\alpha}A_{\beta}(\phi^{\ast}\partial_{\mu}\phi - \partial_{\mu}\phi^{\ast}\phi  ) + \cdots,     
\end{eqnarray} where we have omitted interaction terms that are not relevant to our analysis.
The first term of the equation above is already present in the usual scalar QED and leads to the vertex
\begin{eqnarray}\label{firstvertex}
   \begin{tikzpicture}[baseline=(b3.base)]
 \begin{feynman}
    \vertex (a1);
    \vertex[right=0.9 of a1] (a2);
    \vertex[right=0.9 of a2] (a3);
    \vertex[below=0.9 of a1] (b1);
    \vertex[right=0.9 of b1] (b2);
    \vertex[right=1.3 of b2] (b3);
    \vertex[below=0.9 of b1] (c1);
    \vertex[right=0.9 of c1] (c2);
    \vertex[right=0.9 of c2] (c3);
    \diagram* {
      (a1) -- [charged scalar, momentum=\(p\)] (b2),
      (b2) -- [photon,] (b3),
      (c1) -- [anti charged scalar, momentum'=\(p'\)] (b2),
      };
  \end{feynman}
 \end{tikzpicture} = ie\eta^{\mu \nu}(p_{\mu} - p'_{\mu}). 
\end{eqnarray} The second term of $\mathcal{L}_{int}$, in turn, will be responsible for the corrections due to Lorentz violation for the scattering processes considered here. The interaction vertex associated with it is given by
\begin{eqnarray}\label{secondvertex}
\begin{tikzpicture}[baseline=(b3.base)]
 \begin{feynman}
    \vertex (a1);
    \vertex[right=0.9 of a1] (a2);
    \vertex[right=0.9 of a2] (a3);
    \vertex[below=0.9 of a1] (b1);
    \vertex[dot, right=0.9 of b1] (b2){};
    \vertex[right=1.3 of b2] (b3);
    \vertex[below=0.9 of b1] (c1);
    \vertex[right=0.9 of c1] (c2);
    \vertex[right=0.9 of c2] (c3);
    \diagram* {
      (a1) -- [charged scalar, momentum=\(p\)] (b2),
      (b2) -- [photon,momentum=\(q\)] (b3),
      (c1) -- [anti charged scalar, momentum'=\(p'\)] (b2),
      };
  \end{feynman}
 \end{tikzpicture} = ig\epsilon^{\mu \nu \alpha \beta}\omega_{\nu}q_{\alpha}(p_{\mu} - p'_{\mu}).
\end{eqnarray} Here is an observation. The vertex above takes into account the creation of a photon from the annihilation of two particles. However, if the photon were also being annihilated, an extra minus sign should be added to expression \eqref{secondvertex}. This is due to the factor $\partial_{\alpha}A_{\beta}$ present in the second term of Eq. \eqref{lagintdoistermos}. 


\subsection{Finite Temperature}

The temperature-dependent Feynman rules for the Lorentz-violating scalar QED model described above directly follow from the expression for the scattering matrix in TFD, that is, Eq. \eqref{matrizshatbeta}. The photon propagator in the non-tilde space $\mathcal{S}$, which is defined as 
\begin{eqnarray}\label{propfotontermaldef1}
\langle\, 0(\beta)\, |\mathcal{T}\left[ A_{\mu}(x)A_{\nu}(y) \right]|\, 0(\beta)\, \rangle ,
\end{eqnarray} with
\begin{eqnarray}\label{campodefotonszertotemp}
 A_{\mu}(x) = \int \frac{d^3k}{(2\pi)^3}\, \frac{1}{\sqrt{2E_{\mathbf{k}}}}\, \sum_{m=0}^{3}\left[ a^m(\mathbf{k})\epsilon^m_{\mu}(k)e^{-ik\cdot x} + a^{m \dagger}(\mathbf{k})\epsilon^{\ast m}_{\mu}(k)e^{ik\cdot x} \right],
 \end{eqnarray} being the usual field solution for photons and $\epsilon^m_{\mu}(k)$ being the respective polarization vector, for exemple, can be diagrammatically read, after some algebra, as 
 \begin{eqnarray}\label{thermalphotonprop}
 \begin{tikzpicture}[baseline=(b2.base)]
 \begin{feynman}
    \vertex (b1) {\(\mu\)};
    \vertex [right= 2.0cm of b1] (b2) {\(\nu\)};
    
    \diagram* {
      (b1) -- [photon,  momentum=\(q\)] (b2), 
    };
  \end{feynman}
 \end{tikzpicture} = \frac{-i\eta_{\mu \nu}}{q^2} - 2\pi\, \eta_{\mu \nu}v_B^2(\mathbf{q},\beta)\delta(q^2).
\end{eqnarray} On the same line, it can be shown that the propagator in the tilde space $\Tilde{\mathcal{S}}$, defined as in Eq. \eqref{propfotontermaldef1} by replacing $A_{\mu}$ with $\Tilde{A}_{\mu}$, can be derived simply by acting directly with the conjugation rules \eqref{rct1}-\eqref{rct4} on Eq. \eqref{thermalphotonprop}. The same applies to the interaction vertices, where, when in the space $\mathcal{S}$, they take the form of those presented in equations \eqref{firstvertex} and \eqref{secondvertex}.


\section{Lorentz-violating Differential cross sections for mesons at finite temperature} 
\label{sec4}

In this section, we will calculate the differential cross section at finite temperature using the TFD formalism for three relevant scattering processes in the context of a scalar QED non-minimally violating Lorentz symmetry. The processes can be characterized as follows:
\begin{eqnarray}\label{mamabarmbmbbar}
    M_a(p) + \Bar{M}_a(p') &\rightarrow& M_b(k) + \Bar{M}_b(k');\\ \label{mamabarmamabar}
     M_a(p) + \Bar{M}_a(p') &\rightarrow& M_a(k) + \Bar{M}_a(k');\\
    M_a(p) + M_a(p') &\rightarrow& M_a(k) + M_a(k');\label{mamamama}
\end{eqnarray} where $M_a$ stands for an $a$-type
meson, different from a $b$-type meson, and $\Bar{M}$ stands for the corresponding antiparticles. 

From now on, the procedure to be followed is quite standard in quantum field theory. First, we draw all the Feynman diagrams corresponding to the scattering of interest. Then, we apply the rules presented in the previous section to write the scattering matrix
\begin{eqnarray}\label{mhatbetambetamenosmtildebeta}
  \hat{\mathcal{M}}(\beta) = \mathcal{M}(\beta)-\Tilde{\mathcal{M}}(\beta),  
\end{eqnarray} appropriately defined in Eq. \eqref{matrizshatbeta}. Having said that, let us examine each of the processes in Eqs. \eqref{mamabarmbmbbar}-\eqref{mamamama} separately.

\subsection{Meson-antimeson of $a$-type into meson-antimeson of $b$-type}

At the tree level, the Feynman diagrams, common to both tilde and non-tilde spaces, for the scattering process represented in Eq. \eqref{mamabarmbmbbar} are 
\begin{eqnarray}\label{graffeyntreelschannel}
\begin{tikzpicture}[baseline=(b3.base)]
 \begin{feynman}
    \vertex (a1);
    \vertex [right=1.0 of a1] (a2);
    \vertex [right=1.0 of a2] (a3);
    \vertex [right=1.0 of a3] (a4);
    \vertex [below=1.0 of a1] (b1);
    \vertex [right=1.0 of b1] (b2);
    \vertex [right=1.0 of b2] (b3);
    \vertex [right=1.0 of b3] (b4);
    \vertex [below=1.0 of b1] (c1);
    \vertex [right=1.0 of c1] (c2);
    \vertex [right=1.0 of c2] (c3);
    \vertex [right=1.0 of c3] (c4);

    \diagram* {
      (a1) -- [anti charged scalar, edge label'=\(p'\)] (b2),
      (b2) -- [photon, edge label=\(q\)] (b3), 
      (b3) -- [anti charged scalar, edge label'=\(k'\)] (a4),
      (c1) -- [charged scalar, edge label=\(p\)] (b2),
      (b3) -- [charged scalar, edge label=\(k\)] (c4),
    };
  \end{feynman}
 \end{tikzpicture}\, 
 \begin{tikzpicture}[baseline=(b3.base)]
 \begin{feynman}
    \vertex (a1);
    \vertex [right=1.0 of a1] (a2);
    \vertex [right=1.0 of a2] (a3);
    \vertex [right=1.0 of a3] (a4);
    \vertex [below=1.0 of a1] (b1);
    \vertex [dot, right=1.0 of b1] (b2){};
    \vertex [right=1.0 of b2] (b3);
    \vertex [right=1.0 of b3] (b4);
    \vertex [below=1.0 of b1] (c1);
    \vertex [right=1.0 of c1] (c2);
    \vertex [right=1.0 of c2] (c3);
    \vertex [right=1.0 of c3] (c4);

    \diagram* {
      (a1) -- [anti charged scalar, edge label'=\(p'\)] (b2),
      (b2) -- [photon, edge label=\(q\)] (b3), 
      (b3) -- [anti charged scalar, edge label'=\(k'\)] (a4),
      (c1) -- [charged scalar, edge label=\(p\)] (b2),
      (b3) -- [charged scalar, edge label=\(k\)] (c4),
    };
  \end{feynman}
 \end{tikzpicture}\, 
 \begin{tikzpicture}[baseline=(b3.base)]
 \begin{feynman}
    \vertex (a1);
    \vertex [right=1.0 of a1] (a2);
    \vertex [right=1.0 of a2] (a3);
    \vertex [right=1.0 of a3] (a4);
    \vertex [below=1.0 of a1] (b1);
    \vertex [right=1.0 of b1] (b2);
    \vertex [dot, right=1.0 of b2] (b3){};
    \vertex [right=1.0 of b3] (b4);
    \vertex [below=1.0 of b1] (c1);
    \vertex [right=1.0 of c1] (c2);
    \vertex [right=1.0 of c2] (c3);
    \vertex [right=1.0 of c3] (c4);

    \diagram* {
      (a1) -- [anti charged scalar, edge label'=\(p'\)] (b2),
      (b2) -- [photon, edge label=\(q\)] (b3), 
      (b3) -- [anti charged scalar, edge label'=\(k'\)] (a4),
      (c1) -- [charged scalar, edge label=\(p\)] (b2),
      (b3) -- [charged scalar, edge label=\(k\)] (c4),
    };
  \end{feynman}
 \end{tikzpicture}\,  \begin{tikzpicture}[baseline=(b3.base)]
 \begin{feynman}
    \vertex (a1);
    \vertex [right=1.0 of a1] (a2);
    \vertex [right=1.0 of a2] (a3);
    \vertex [right=1.0 of a3] (a4);
    \vertex [below=1.0 of a1] (b1);
    \vertex [dot, right=1.0 of b1] (b2){};
    \vertex [dot, right=1.0 of b2] (b3){};
    \vertex [right=1.0 of b3] (b4);
    \vertex [below=1.0 of b1] (c1);
    \vertex [right=1.0 of c1] (c2);
    \vertex [right=1.0 of c2] (c3);
    \vertex [right=1.0 of c3] (c4);

    \diagram* {
      (a1) -- [anti charged scalar, edge label'=\(p'\)] (b2),
      (b2) -- [photon, edge label=\(q\)] (b3), 
      (b3) -- [anti charged scalar, edge label'=\(k'\)] (a4),
      (c1) -- [charged scalar, edge label=\(p\)] (b2),
      (b3) -- [charged scalar, edge label=\(k\)] (c4),
    };
  \end{feynman}
 \end{tikzpicture},
\end{eqnarray} where we have considered up to second order in the Lorentz symmetry violation parameter. By means of the Feynman rules discussed in the previous section, we can express Eq. \eqref{mhatbetambetamenosmtildebeta} as
\begin{eqnarray}\label{mhatbetatotal30}
    \hat{\mathcal{M}}(\beta) &=& \frac{1}{q^2} \left(\prod v_B\right)\left(1+\prod \frac{u_B}{v_B}\right)\left\lbrace 1+2 \pi i\,  v_B^2(\mathbf{q},\beta)\, q^2\, \delta(q^2)\left[ \frac{1-\prod \frac{u_B}{v_B}}{1+\prod \frac{u_B}{v_B}} \right]\right\rbrace  \nonumber\\
    &\times& \left\lbrace e^2 \left( k'\cdot p' - p\cdot k'-k\cdot p'+k\cdot p\right) + 2eg\, \epsilon^{\mu \nu \alpha \beta}\, \omega_{\nu}\, q_{\alpha}\, (k_{\beta}-k'_{\beta})(p_{\mu}-p'_{\mu}) \right.\nonumber\\
    &+& \left. g^2\, \epsilon^{\mu \nu \alpha \beta}\, \epsilon^{\rho \sigma \gamma \eta}\, \eta_{\beta \rho}\, \omega_{\nu}\, \omega_{\gamma}\, q_{\alpha}\, q_{\eta}\, (k_{\sigma}-k'_{\sigma})(p_{\mu}-p'_{\mu})\right\rbrace,
\end{eqnarray} where 
\begin{eqnarray}
    \prod v_B \equiv v_B(\vec{p},\beta)\, v_B(\vec{p'},\beta)\, v_B(\vec{k},\beta)\, v_B(\vec{k'},\beta), 
\end{eqnarray} and
\begin{eqnarray}
    \prod \frac{u_B}{v_B} \equiv \frac{u_B(\vec{p},\beta)\, u_B(\vec{p'},\beta)\, u_B(\vec{k},\beta)\, u_B(\vec{k'},\beta)}{v_B(\vec{p},\beta)\, v_B(\vec{p'},\beta)\, v_B(\vec{k},\beta)\, v_B(\vec{k'},\beta)}.
\end{eqnarray} Note that $\mathcal{\hat{M}}=\mathcal{\hat{M}}_s$, where the subscript `$s$' indicates that we are only dealing with s-channel diagrams. From Eq. \eqref{mhatbetatotal30}, we can directly compute $|\hat{\mathcal{M}}(\beta)|^2$ and evaluate the result from the center-of-mass reference frame of the system, where 
\begin{eqnarray}\label{setcentermassframe33}
p=(E,E\, \hat{\mathbf{z}}); &\hspace{1cm}& k=(E,\mathbf{k}); \nonumber\\
p'=(E,-E\, \hat{\mathbf{z}}); &\hspace{1cm}&  k'=(E,-\mathbf{k}); 
\end{eqnarray} with
\begin{eqnarray}
 \hat{\mathbf{x}}\cdot \mathbf{k} &=& E\sin \theta \cos \phi,\\ 
 \hat{\mathbf{y}}\cdot \mathbf{k} &=& E\sin \theta \sin \phi,\\
 \hat{\mathbf{z}}\cdot \mathbf{k} &=& E\cos\theta,
\end{eqnarray} and $E_{cm}=2E$. This evaluation allows us to calculate the cross section using Eq. \eqref{secdifbetacmsomaspinmhatcmbetaquadrado}. Note that when writing Eqs. \eqref{setcentermassframe33}, we are implicitly assuming that our analysis is conducted in the high-energy limit, where we can take $m_a=m_b=0$, with $m_a$ and $m_b$ representing the masses of the particles of type $a$ and $b$, respectively. In addition, concerning the Lorentz violation parameter, we will separately investigate two distinct possibilities based on whether $\omega^{\mu}$ is a timelike or spacelike four-vector.

\subsubsection{Timelike case}

\begin{figure}
\includegraphics[scale=0.74]{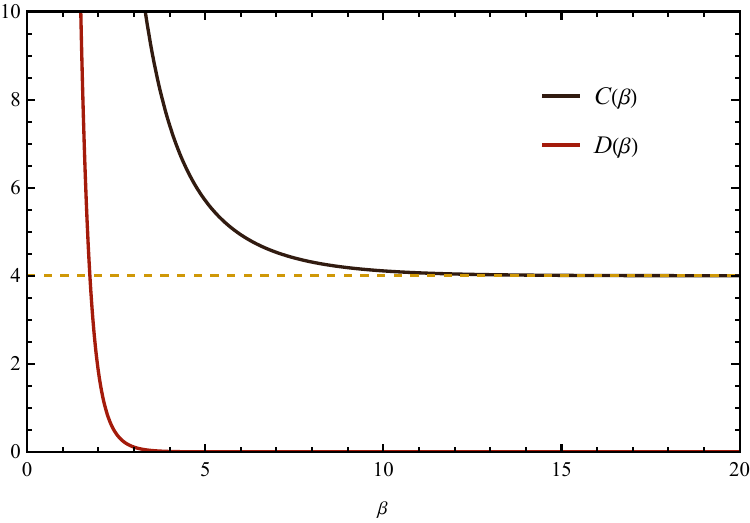}
    \caption{\raggedright Graph of temperature-dependent functions $C$ and $D$ as a function of $\beta$. We have considered $E_{cm}=1$.}
    \label{canddgraph}
\end{figure}

In this case, we must consider $\omega^{\nu} = (\omega_0,\mathbf{0})$, where $\omega_0$ is a constant parameter that sets the Lorentz violation effects. In accordance with what was previously established, after some algebraic manipulation, we can write 
\begin{eqnarray}
    |\hat{\mathcal{M}}(\beta)|^2 = B(\beta)\, e^4\, \cos^2 \theta,
\end{eqnarray} where
\begin{eqnarray}\label{bdebetaecmdeltaprodargquad}
  B(\beta) = \frac{1}{4}\left\lbrace C(\beta)+\pi^2\, \delta \left(E_{cm}^2\right)^2\, D(\beta) \right\rbrace,   
\end{eqnarray}
is the thermal correction factor for a meson and antimeson of an $a$-type scattering off a meson and antimeson of a $b$-type. The temperature-dependent functions $C(\beta)$ and $D(\beta)$ are given by  
\begin{eqnarray}\label{cfunctionbetta}
    C(\beta)= \cosh^2 \left( \frac{\beta\, E_{cm}}{2} \right) \text{csch}^4\left(\frac{\beta\, E_{cm}}{4}\right),
\end{eqnarray} and 
\begin{eqnarray}\label{dfunctionbetta}
    D(\beta)= \left( \frac{2\, E_{cm}}{e^{\frac{\beta E_{cm}}{2}}-1} \right)^4,
\end{eqnarray} respectively. Figure \eqref{canddgraph} illustrates the behavior of these functions in terms of the parameter $\beta$. 

The temperature-dependent differential cross section for this specific configuration can be written from Eq. \eqref{secdifbetacmsomaspinmhatcmbetaquadrado} as
\begin{eqnarray}
    \left( \frac{d\sigma}{d\Omega} \right)_{\beta,cm} &=& B(\beta) \left( \frac{d\sigma_0}{d\Omega} \right)_{cm}
\end{eqnarray} where
\begin{eqnarray}\label{secdifzerotempnoviolation}
    \left( \frac{d\sigma_0}{d\Omega} \right)_{cm} = \frac{e^4}{64\, \pi^2\, E_{cm}^2}\cos^2 \theta,
\end{eqnarray} represents the corresponding differential cross section at zero temperature.

The above result reveals that, although the cross section is corrected for temperature through a factor $B(\beta)$, no Lorentz violation effect will be observed if the vector $\omega^{\nu}$ characterizing it is timelike. In fact, the expression in Eq. \eqref{secdifzerotempnoviolation} represents the usual outcome for the scattering of distinct mesons at zero temperature, thus aligning completely with what was found in reference \cite{Araujo:2022qke}.

\begin{figure}[!]
\centering
\includegraphics[scale=0.6]{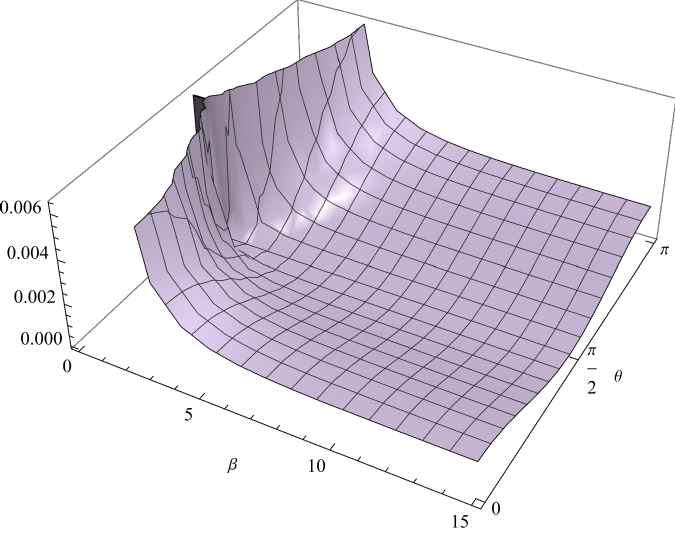}
\caption{\raggedright Angular and temperature dependencies of the differential cross section for $a$-type meson-antimeson scattering in $b$-type meson-antimeson in the timelike case. For this plot we have considered $e=1$ and  $E_{cm} = 1$.}\label{difsectime1}
\end{figure}

In terms of the dependence of the cross section on temperature, we observe that it is entirely governed by the function in Eq. \eqref{bdebetaecmdeltaprodargquad}. As we can notice (see Figure \eqref{canddgraph}), the temperature corrections become highly relevant in the limit of very high temperatures, where $C(\beta)$ and $D(\beta)$, and consequently $B(\beta)$, assumes large values. Conversely, as the temperature approaches zero, Eq. \eqref{bdebetaecmdeltaprodargquad} tends towards unity, allowing us to recover the usual result. Figure \eqref{difsectime1} shows the behavior of the differential thermal cross section in terms of the temperature and the angle $\theta$. As we can also observe, there still exists in $B(\beta)$ a product of delta function with equal arguments, namely $ \delta \left(E_{cm}^2\right)^2$. Such a term is highly characteristic in the TFD formalism and represents a kind of apparent "pathology" that are actually called pinch singularity \cite{Landsman:1986uw}. However, as mentioned in Refs. \cite{Landsman:1986uw,van2001introduction}, all these problems are avoided by working with the regularized form of delta-functions
and their derivatives, namely
\begin{eqnarray}\label{propriedadedaderivadadadeltaprop}
 2\pi i \frac{1}{n!}\frac{\partial^n}{\partial x^n }\delta (x) = \left( -\frac{1}{x+i\epsilon} \right)^{n+1} - \left( -\frac{1}{x-i\epsilon} \right)^{n+1}.
\end{eqnarray} 

\subsubsection{Spacelike case}

\begin{figure}[!]
     \centering
     \begin{subfigure}[b]{0.3\textwidth}
         \centering
         \includegraphics[scale=0.6]{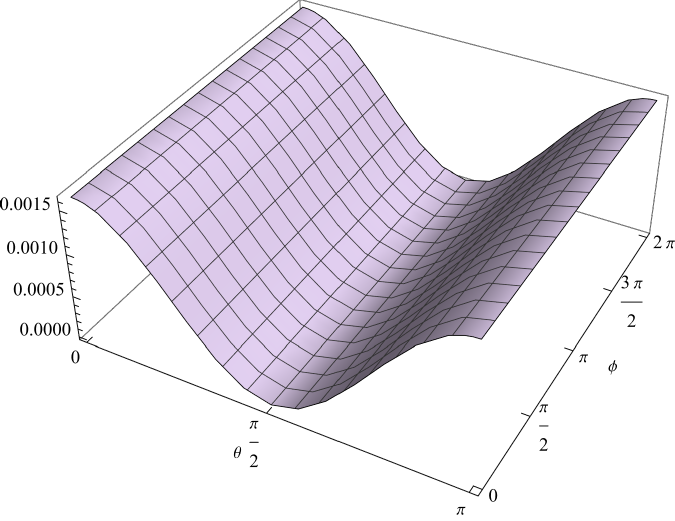}
         \caption{}
         \label{difsecspace1a}
     \end{subfigure}
     \hfill
     \begin{subfigure}[b]{0.6\textwidth}
         \centering
         \includegraphics[scale=0.6]{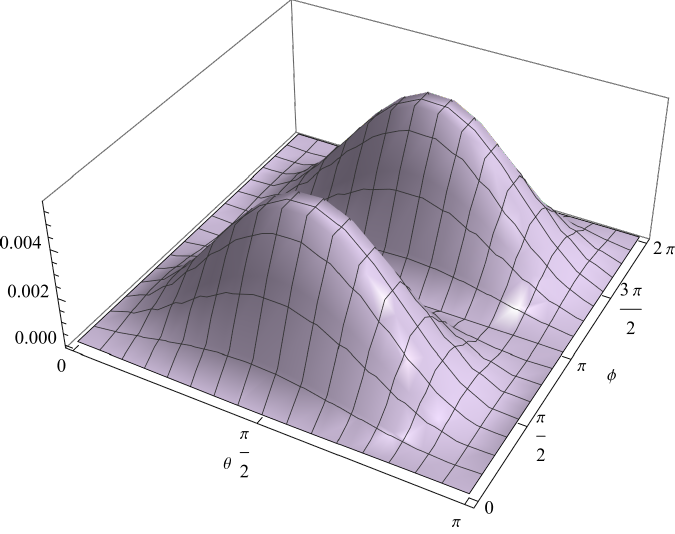}
         \caption{}
         \label{difsecspace1b}
     \end{subfigure}
     \hfill
     \begin{subfigure}[b]{0.6\textwidth}
         \centering
         \includegraphics[scale=0.6]{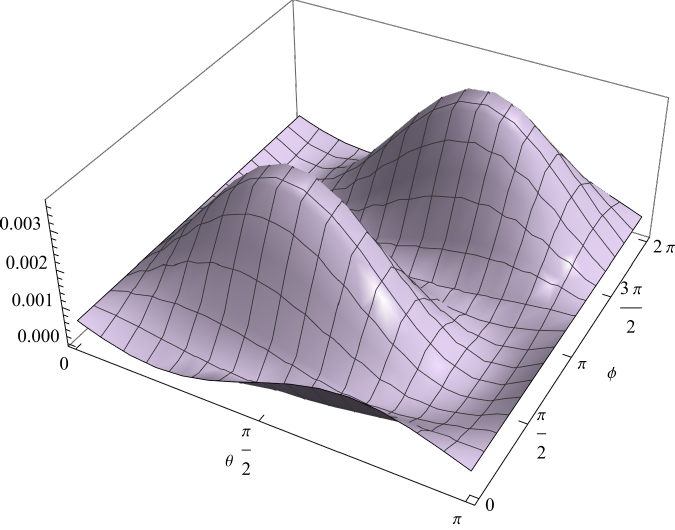}
         \caption{}
         \label{difsecspace1c}
     \end{subfigure}
        \caption{\raggedright Angular dependencies of the zero-temperature differential cross section for $a$-type meson-antimeson scattering in $b$-type meson-antimeson in the spacelike case. For this plot we took $e=1$,  $E_{cm} = 1$. In \eqref{difsecspace1a}, we have the usual case ($g|\mathbf{w}| =0$). In \eqref{difsecspace1b} and \eqref{difsecspace1c}, we have considered $g|\mathbf{w}| =0.9$, but for the former $\theta_{\omega z}=\frac{\pi}{2}$ and $\phi_{\omega x}=0$, while for the latter $\theta_{\omega z}=\phi_{\omega x}=\frac{\pi}{4}$.}
        \label{fig3space}
\end{figure}

For this case, we must consider $\omega^{\nu}=(0,\mathbf{w})$, where $\mathbf{w}$ is a constant vector in an arbitrary space direction. Let $\theta_{\omega z}$ be the angle that the vector $\mathbf{w}$ makes with the z-axis, and $\phi_{\omega x}$ the angle that the projection of $\mathbf{w}$ onto the xy-plane makes with the x-axis. Then, following the same steps as in the previous case, after some algebra, we can express the temperature-dependent differential cross-section as 
\begin{eqnarray}\label{difsecbetacmbdebetazerdisec}
 \left( \frac{d\sigma}{d\Omega} \right)_{\beta,cm} &=& B(\beta) \left( \frac{d\sigma}{d\Omega} \right)_{cm},   
\end{eqnarray} where 
\begin{eqnarray}\label{difcrosssecspacelikea2}
    \left( \frac{d\sigma}{d\Omega} \right)_{cm} &=& \frac{e^4 }{64\, \pi ^2\, E_{cm}^2}\cos ^2\theta  + g\, |\mathbf{w}|\, \frac{e^3}{32\, \pi ^2\,
   E_{cm}}\, \sin (2 \theta) \, \sin (\theta _{\omega z})\, \sin \left(\phi -\phi _{\omega x}\right)\nonumber\\
   &+& g^2\, |\mathbf{w}|^2 \, \frac{e^2}{32\, \pi ^2}\, \sin ^2 (\theta _{\omega z}) \Big[\sin (\theta) \, \cos (\theta) \, \cot (\theta _{\omega
   z})\, \cos \left(\phi -\phi _{\omega x}\right)  \nonumber\\ 
   &+&  2 \sin ^2(\theta) \, \sin ^2\left(\phi -\phi _{\omega x}\right)-\cos
   ^2(\theta) \Big]\nonumber\\
   &-& g^3\, |\mathbf{w}|^3\, \frac{e\, E_{cm}}{16\, \pi ^2}\, \sin ^2(\theta) \, \sin ^3 (\theta _{\omega z})\, \sin \left(\phi -\phi _{\omega x}\right)
   \Big[\cot (\theta) -\cot (\theta _{\omega z})\, \cos \left(\phi -\phi _{\omega x}\right)\Big]\nonumber\\
   &-& g^4\, |\mathbf{w}|^4\, \frac{E_{cm}^2}{1024\, \pi ^2}\, \sin ^2(\theta _{\omega z})\, \Big\lbrace \cos (2 \theta) \, \big[ 2 \cos (2 \theta _{\omega z}) \big( 3 + \cos (2 \left(\phi -\phi _{\omega x}\right)) \big)+1 \big] \nonumber\\
   &+& 4 \cos (2 \theta _{\omega z})\,  \sin ^2\left(\phi -\phi _{\omega x}\right)+8 \sin (2 \theta) \, \sin (2 \theta _{\omega z})\, \cos
   \left(\phi -\phi _{\omega x}\right)\nonumber\\
   &+& 2 \sin ^2(\theta) \, \cos (2 \left(\phi -\phi _{\omega x}\right))-9 \Big\rbrace ,
\end{eqnarray} corresponds to the result at zero temperature, taking into account terms up to fourth order in the Lorentz violation parameter. From Eq. \eqref{difsecbetacmbdebetazerdisec}, we can observe once again that the behavior of the differential cross section with temperature is entirely governed by the function $B(\beta)$, such that corrections become highly relevant in the limit of very high temperatures, and the usual result ($T=0$) is promptly recovered. However, unlike the timelike case, Lorentz violation effects are now observed, suggesting that the scattering in question is more sensitive in scenarios where $\omega^{\mu}$ is a spacelike four-vector. Note also that such effects are completely canceled for situations where $\mathbf{w}$ lies in the plane containing the line defining the collision trajectory of the particles at the beginning of the process. Figure \eqref{fig3space} illustrates the angular dependencies of the differential cross section at zero temperature for some values of $\theta_{\omega z}$ and $\phi_{\omega x}$. It is evident that the behavior changes dramatically from one situation to another, particularly concerning the region where the differential cross section reaches its maximum value.

\subsection{Meson-antimeson of $a$-type into meson-antimeson of $a$-type}\label{secmamatypemamatype}

When considering the scattering process represented by Eq. \eqref{mamabarmamabar} at the tree level, we must also take into account four more Feynman diagrams, namely 
\begin{eqnarray}\label{graphmamammama12345}
\begin{tikzpicture}[baseline=(b3.base)]
 \begin{feynman}
    \vertex (a1);
    \vertex [below=1.0 of a1] (a2);
    \vertex [below=1.0 of a2] (a3);
    \vertex [below=1.0 of a3] (a4);
    \vertex [right=1.0 of a1] (b1);
    \vertex [below=1.0 of b1] (b2);
    \vertex [below=1.0 of b2] (b3);
    \vertex [below=1.0 of b3] (b4);
    \vertex [right=1.0 of b1] (c1);
    \vertex [below=1.0 of c1] (c2);
    \vertex [below=1.0 of c2] (c3);
    \vertex [below=1.0 of c3] (c4);

    \diagram* {
      (a1) -- [anti charged scalar, edge label'=\(p'\)] (b2),
      (b2) -- [photon, edge label=\(q'\)] (b3), 
      (b3) -- [anti charged scalar, edge label'=\(p\)] (a4),
      (c1) -- [charged scalar, edge label=\(k'\)] (b2),
      (b3) -- [charged scalar, edge label=\(k\)] (c4),
    };
  \end{feynman}
 \end{tikzpicture}\, \begin{tikzpicture}[baseline=(b3.base)]
 \begin{feynman}
    \vertex (a1);
    \vertex [below=1.0 of a1] (a2);
    \vertex [below=1.0 of a2] (a3);
    \vertex [below=1.0 of a3] (a4);
    \vertex [right=1.0 of a1] (b1);
    \vertex [dot, below=1.0 of b1] (b2){};
    \vertex [below=1.0 of b2] (b3);
    \vertex [below=1.0 of b3] (b4);
    \vertex [right=1.0 of b1] (c1);
    \vertex [below=1.0 of c1] (c2);
    \vertex [below=1.0 of c2] (c3);
    \vertex [below=1.0 of c3] (c4);

    \diagram* {
      (a1) -- [anti charged scalar, edge label'=\(p'\)] (b2),
      (b2) -- [photon, edge label=\(q'\)] (b3), 
      (b3) -- [anti charged scalar, edge label'=\(p\)] (a4),
      (c1) -- [charged scalar, edge label=\(k'\)] (b2),
      (b3) -- [charged scalar, edge label=\(k\)] (c4),
    };
  \end{feynman}
 \end{tikzpicture}\, \begin{tikzpicture}[baseline=(b3.base)]
 \begin{feynman}
    \vertex (a1);
    \vertex [below=1.0 of a1] (a2);
    \vertex [below=1.0 of a2] (a3);
    \vertex [below=1.0 of a3] (a4);
    \vertex [right=1.0 of a1] (b1);
    \vertex [below=1.0 of b1] (b2);
    \vertex [dot, below=1.0 of b2] (b3){};
    \vertex [below=1.0 of b3] (b4);
    \vertex [right=1.0 of b1] (c1);
    \vertex [below=1.0 of c1] (c2);
    \vertex [below=1.0 of c2] (c3);
    \vertex [below=1.0 of c3] (c4);

    \diagram* {
      (a1) -- [anti charged scalar, edge label'=\(p'\)] (b2),
      (b2) -- [photon, edge label=\(q'\)] (b3), 
      (b3) -- [anti charged scalar, edge label'=\(p\)] (a4),
      (c1) -- [charged scalar, edge label=\(k'\)] (b2),
      (b3) -- [charged scalar, edge label=\(k\)] (c4),
    };
  \end{feynman}
 \end{tikzpicture}\, \begin{tikzpicture}[baseline=(b3.base)]
 \begin{feynman}
    \vertex (a1);
    \vertex [below=1.0 of a1] (a2);
    \vertex [below=1.0 of a2] (a3);
    \vertex [below=1.0 of a3] (a4);
    \vertex [right=1.0 of a1] (b1);
    \vertex [dot, below=1.0 of b1] (b2){};
    \vertex [dot, below=1.0 of b2] (b3){};
    \vertex [below=1.0 of b3] (b4);
    \vertex [right=1.0 of b1] (c1);
    \vertex [below=1.0 of c1] (c2);
    \vertex [below=1.0 of c2] (c3);
    \vertex [below=1.0 of c3] (c4);

    \diagram* {
      (a1) -- [anti charged scalar, edge label'=\(p'\)] (b2),
      (b2) -- [photon, edge label=\(q'\)] (b3), 
      (b3) -- [anti charged scalar, edge label'=\(p\)] (a4),
      (c1) -- [charged scalar, edge label=\(k'\)] (b2),
      (b3) -- [charged scalar, edge label=\(k\)] (c4),
    };
  \end{feynman}
 \end{tikzpicture}\, ,
\end{eqnarray} in addition to those depicted in Eq. \eqref{graffeyntreelschannel}. These new diagrams correspond to the t-channel of the interaction process in question, such that the total thermal scattering matrix $\hat{\mathcal{M}}(\beta)$ now becomes  
\begin{eqnarray}
    \hat{\mathcal{M}}(\beta) = \hat{\mathcal{M}}_s(\beta) + \hat{\mathcal{M}}_t(\beta),
\end{eqnarray} with  
\begin{eqnarray}\label{matrixchannelthatbeta}
    \hat{\mathcal{M}}_t(\beta) &=& \frac{e^2}{q'^{2}} \left(\prod v_B\right)\left(1+\prod \frac{u_B}{v_B}\right)\left\lbrace 1+2 \pi i\,  v_B^2(\mathbf{q'},\beta)\, q'^2\, \delta(q'^2)\left[ \frac{1-\prod \frac{u_B}{v_B}}{1+\prod \frac{u_B}{v_B}} \right]\right\rbrace  \nonumber\\
    &\times& \left\lbrace  -\left( p\cdot k' + k\cdot k'+k\cdot p'+p\cdot p'\right) + 2\frac{g}{e}\, \epsilon^{\mu \nu \alpha \beta}\, \omega_{\nu}\, q'_{\alpha}\, (k_{\beta}+p_{\beta})(k'_{\mu}+p'_{\mu}) \right.\nonumber\\
    &+& \left. \frac{g^2}{e^2}\, \epsilon^{\mu \nu \alpha \beta}\, \epsilon^{\sigma \gamma \eta \rho}\, \eta_{\beta \rho}\, \omega_{\nu}\, \omega_{\gamma}\, q'_{\alpha}\, q'_{\eta}\, (k'_{\sigma}+p'_{\sigma})(k_{\mu}+p_{\mu})\right\rbrace,
\end{eqnarray}
representing the contribution from the $t$-channel and $\hat{\mathcal{M}}_s(\beta)$ (still given by Eq. \eqref{mhatbetatotal30}) the contribution from the $s$-channel, both constructed according to the Feynman rules discussed in section \eqref{sec3}. As before, by knowing $\hat{\mathcal{M}}(\beta)$, we can evaluate its squared matrix in the center-of-mass reference frame (see Eq. \eqref{setcentermassframe33}) and write, through Eq. \eqref{secdifbetacmsomaspinmhatcmbetaquadrado}, an expression for the thermal differential cross section in the two cases of interest: when $\omega^{\mu}$ is a timelike four-vector and when $\omega^{\mu}$ is a spacelike four-vector. 

\subsubsection{Timelike case}

\begin{figure}[!]
     \centering
     \begin{subfigure}[b]{0.3\textwidth}
         \centering
         \includegraphics[scale=0.6]{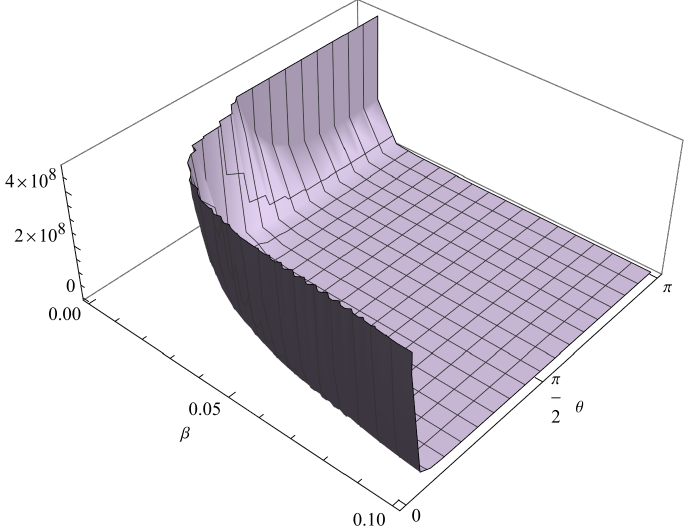}
         \caption{}
         \label{difsectime2a}
     \end{subfigure}
     \hfill
     \begin{subfigure}[b]{0.6\textwidth}
         \centering
         \includegraphics[scale=0.6]{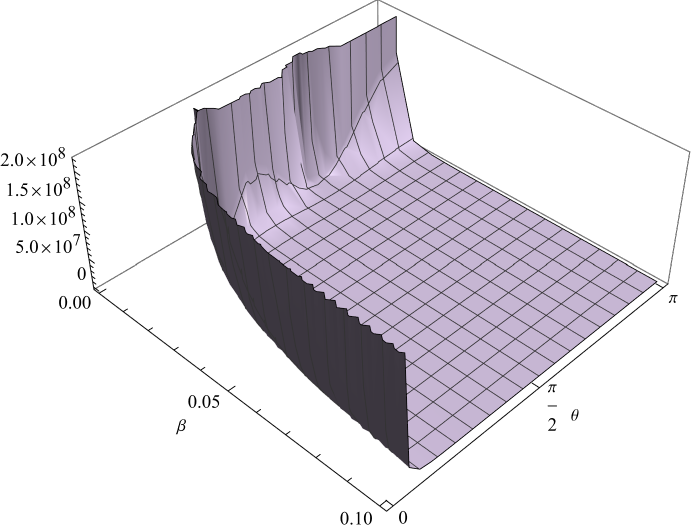}
         \caption{}
         \label{difsectime2b}
     \end{subfigure}
        \caption{\raggedright Angular and temperature dependencies of the differential cross section for $a$-type meson-antimeson scattering in $a$-type meson-antimeson in the timelike case. For this plot we took $e=1$ and  $E_{cm} = 1$. In \eqref{difsectime2a}, no Lorentz violation has been considered ($g \omega_0=0$). In \eqref{difsectime2b}, we have set $g \omega_0=3$. }
        \label{fig4timecomsemlv}
\end{figure}

Since $\omega^{\nu} = (\omega_0,\mathbf{0})$ for this case, after some algebraic manipulation, the temperature-dependent differential cross section for the scattering process characterized in Eq. \eqref{mamabarmamabar} takes the form 
\begin{eqnarray}\label{difcrosectimemamamama}
   \left( \frac{d\sigma}{d\Omega} \right)_{\beta,cm} &=& \frac{e^4}{4096\, \pi^2 \, E_{cm}^2} \Biggl\{ (7+\cos (2\theta))^2 \csc^4\left(\frac{\theta}{2}\right)C(\beta) \nonumber\\
    &+&  16\, \pi^2\, D(\beta)\Big[ \cos (\theta) \, \delta \left(E_{cm}^2\right) + \left( 3+ \cos (\theta) \right)\delta \left( E_{cm}^2(\cos (\theta)-1) \right) \Big]^2 \Biggr\} \nonumber\\
    &-& \frac{1}{512}\left( \frac{e\, g\, \omega_0}{\pi} \right)^2 \Biggl\{  (7+\cos (2\theta))\cot^2\left(\frac{\theta}{2}\right)C(\beta) \nonumber\\
    &+& \, \pi^2\, D(\beta) \sin (\theta)\, \delta \left(E_{cm}^2(\cos (\theta) -1)\right)\big[ \sin (2\theta)\, \delta \left(E_{cm}^2\right) \nonumber\\
    &+& \left( 6 \sin (\theta) + \sin (2\theta) \right) \delta \left(E_{cm}^2(\cos (\theta) -1)\right)  \big] \Biggr\}\nonumber\\
    &+& \frac{\pi^2\, E_{cm}^2}{256}\left( \frac{g\, \omega_0}{\pi} \right)^4 \Biggl\{ \cos^4\left(\frac{\theta}{2}\right)C(\beta)\nonumber\\
    &+& \frac{\pi^2}{4}D(\beta)\sin^4(\theta)\Big[\delta \left(E_{cm}^2(\cos \theta -1)\right)\Big]^2 \Biggr\},
\end{eqnarray}  where $C(\beta)$ and $D(\beta)$ are again the temperature-dependent functions given respectively in Eqs. \eqref{cfunctionbetta} and \eqref{dfunctionbetta}. Note that, in comparison with what was found in the previous case, the result in Eq. \eqref{difcrosectimemamamama} suggests that the scattering process described by Eq. \eqref{mamabarmamabar} is more sensitive to Lorentz violation effects when these are characterized by the parameter $\omega_0$ of the timelike four-vector $\omega^{\mu}$. However, observe that neither first-order nor third-order terms in the coupling constant $g$ are present in the above result. As we will see, this seems to be a characteristic of scattering involving mesons and antimesons of the same species within the context of the non-minimally Lorentz-violating scalar QED framework presented in this work. Figure \eqref{fig4timecomsemlv} illustrates the behavior of the differential cross section with the temperature ($\beta=1/T$) and the angle $\theta$ for the cases with and without Lorentz symmetry violation. As we can observe, the effects of anisotropy characterized by the parameter $\omega_0$ become more pronounced in the limit of very high temperatures, suggesting a new perspective (or path) in the search for Lorentz-violating measurable effects.

\subsubsection{Spacelike case}

\begin{figure}[!]
     \centering
     \begin{subfigure}[b]{0.3\textwidth}
         \centering
         \includegraphics[scale=0.6]{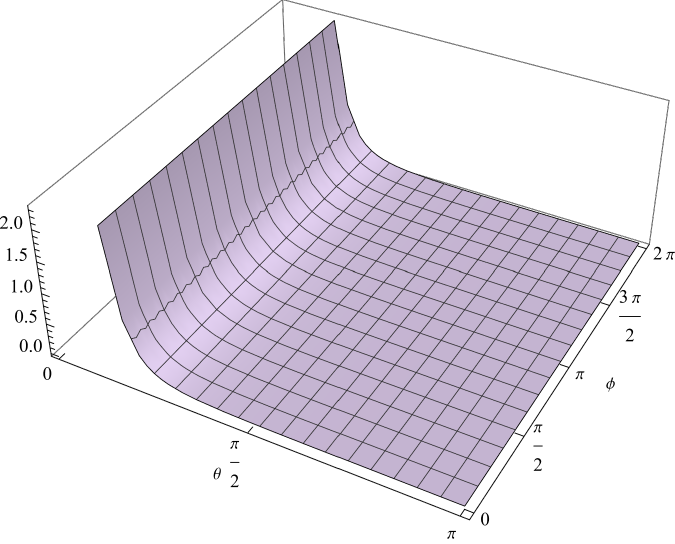}
         \caption{}
         \label{difsecspace2a}
     \end{subfigure}
     \hfill
     \begin{subfigure}[b]{0.6\textwidth}
         \centering
         \includegraphics[scale=0.6]{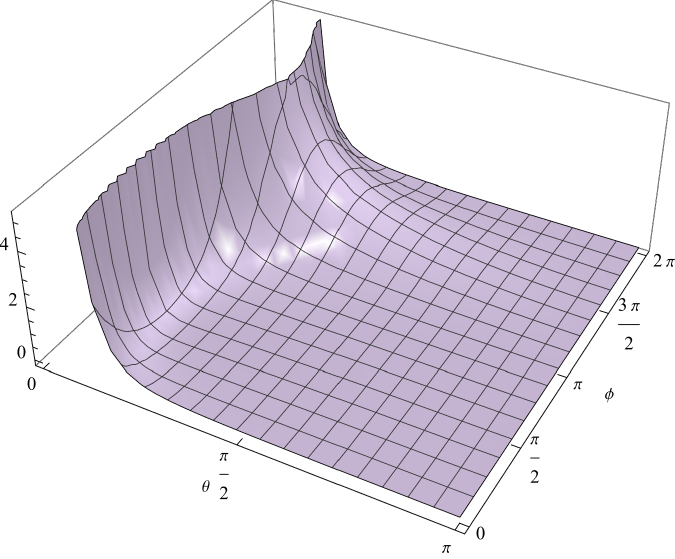}
         \caption{}
         \label{difsecspace2b}
     \end{subfigure}
     \hfill
     \begin{subfigure}[b]{0.6\textwidth}
         \centering
         \includegraphics[scale=0.6]{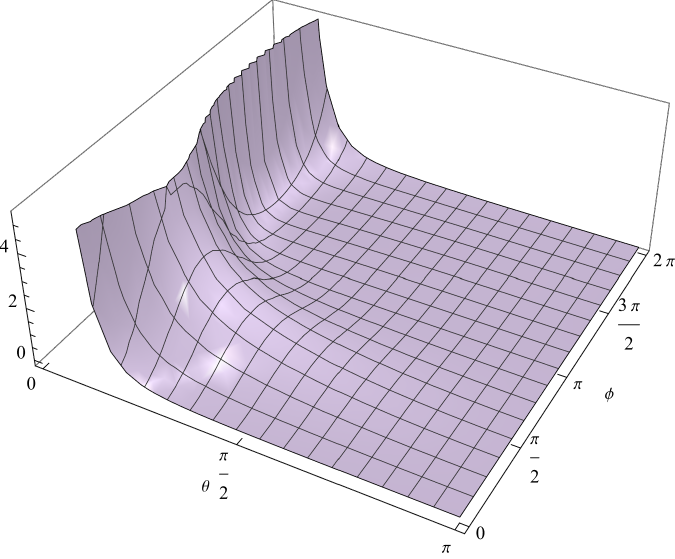}
         \caption{}
         \label{difsecspace2c}
     \end{subfigure}
        \caption{\raggedright Angular dependencies of the zero-temperature differential cross section for $a$-type meson-antimeson scattering in $a$-type meson-antimeson in the spacelike case. For this plot we took $e=1$,  $E_{cm} = 1$. In \eqref{difsecspace2a}, we have the usual case ($g|\mathbf{w}| =0$). In \eqref{difsecspace2b} and \eqref{difsecspace2c}, we have considered $g|\mathbf{w}| =2$, but for the former $\theta_{\omega z}=\frac{\pi}{2}$ and $\phi_{\omega x}=0$, while for the latter $\theta_{\omega z}=\frac{\pi}{2}$ and $\phi_{\omega x}=\pi$.}
        \label{fig5difsecspaceaaaabbb}
\end{figure}

For this case, we must again consider that $\omega^{\nu}=(0,\mathbf{w})$, with $\mathbf{w}$ being the constant vector that establishes the effects of Lorentz violation. Consequently, proceeding exactly as before, the temperature-dependent differential cross section takes the form 

\begin{align}\label{difcrosssecspacelikeb2}
     \left( \frac{d\sigma}{d\Omega} \right)_{\beta,cm} &= \frac{e^4}{4096\, \pi^2 \, E_{cm}^2} \Biggl\{ (7+\cos (2\theta))^2 \csc^4\left(\frac{\theta}{2}\right)C(\beta) \nonumber\\
    &+  16\, \pi^2\, D(\beta)\Big[ \cos (\theta) \, \delta \left(E_{cm}^2\right) + \left( 3+ \cos (\theta) \right)\delta \left( E_{cm}^2(\cos{\theta}-1) \right) \Big]^2 \Biggr\} \nonumber\\
    &+ g\,  |\mathbf{w}|\, \frac{e^3}{256\, \pi^2\, E_{cm}}\nonumber\\
    &\times \Biggl\{(\cos (\theta) -3)(7+\cos (2\theta))\cot \left(\frac{\theta}{2}\right )\csc^2 \left( \frac{\theta}{2} \right)\sin (\theta_{\omega z})\sin (\phi-\phi_{\omega x})\, C(\beta)\nonumber\\
    &+ 4\pi^2\, D(\beta)\, \sin (\theta)\,  \sin (\theta_{\omega z})\, \sin (\phi-\phi_{\omega x})\, \Big[\,  \delta \left(E_{cm}^2\right)-2\, \delta \left( E_{cm}^2(\cos (\theta) -1) \right)\Big]\nonumber\\
    &\times \Big[ \cos (\theta) \, \delta \left(E_{cm}^2\right) + \left( 3+ \cos (\theta) \right)\delta \left( E_{cm}^2(\cos{\theta}-1) \right) \Big]\Bigg\}
   + \cdots,
\end{align} where the dots stands for higher-orders Lorentz-breaking terms. The exact form of these corrections are
lengthy and will not be displayed in detail here. However, we must emphasize that all the terms up to fourth order in $g$ are once again present in the above result. This suggests that the scattering processes involving mesons and antimesons of the same type are more sensitive to Lorentz violation effects characterized by a spacelike four-vector within the context of the QED framework here discussed. From Eq. \eqref{difcrosssecspacelikeb2}, we again see that the temperature dependence is shaped by functions $C(\beta)$ and $D(\beta)$, indicating that at very high temperatures, thermal effects cannot be neglected. Figure \eqref{fig5difsecspaceaaaabbb} illustrates the angular dependencies of the differential cross section at zero temperature for some values of $\theta_{\omega z}$ and $\phi_{\omega x}$. It is noteworthy that the effects of anisotropy cancel out for $\theta=\pi$, irrespective of the azimuthal angle $\phi$ and the orientation of $\mathbf{w}$.

\subsection{Meson-meson of $a$-type into meson-meson of $a$-type}

Characterized by Eq. \eqref{mamamama}, the scattering process involving only mesons of the same type should be evaluated at the tree level in terms of six Feynman diagrams, namely
\begin{eqnarray}\label{graphmmamma111222}
 &\begin{tikzpicture}[baseline=(b3.base)]
 \begin{feynman}
    \vertex (a1);
    \vertex [below=1.0 of a1] (a2);
    \vertex [below=1.0 of a2] (a3);
    \vertex [below=1.0 of a3] (a4);
    \vertex [right=1.0 of a1] (b1);
    \vertex [below=1.0 of b1] (b2);
    \vertex [below=1.0 of b2] (b3);
    \vertex [below=1.0 of b3] (b4);
    \vertex [right=1.0 of b1] (c1);
    \vertex [below=1.0 of c1] (c2);
    \vertex [below=1.0 of c2] (c3);
    \vertex [below=1.0 of c3] (c4);

    \diagram* {
      (a1) -- [charged scalar, edge label'=\(p'\)] (b2),
      (b2) -- [photon, edge label=\(q'\)] (b3), 
      (b3) -- [anti charged scalar, edge label'=\(p\)] (a4),
      (c1) -- [anti charged scalar, edge label=\(k'\)] (b2),
      (b3) -- [charged scalar, edge label=\(k\)] (c4),
    };
  \end{feynman}
 \end{tikzpicture}\, \begin{tikzpicture}[baseline=(b3.base)]
 \begin{feynman}
    \vertex (a1);
    \vertex [below=1.0 of a1] (a2);
    \vertex [below=1.0 of a2] (a3);
    \vertex [below=1.0 of a3] (a4);
    \vertex [right=1.0 of a1] (b1);
    \vertex [dot, below=1.0 of b1] (b2){};
    \vertex [below=1.0 of b2] (b3);
    \vertex [below=1.0 of b3] (b4);
    \vertex [right=1.0 of b1] (c1);
    \vertex [below=1.0 of c1] (c2);
    \vertex [below=1.0 of c2] (c3);
    \vertex [below=1.0 of c3] (c4);

    \diagram* {
      (a1) -- [charged scalar, edge label'=\(p'\)] (b2),
      (b2) -- [photon, edge label=\(q'\)] (b3), 
      (b3) -- [anti charged scalar, edge label'=\(p\)] (a4),
      (c1) -- [anti charged scalar, edge label=\(k'\)] (b2),
      (b3) -- [charged scalar, edge label=\(k\)] (c4),
    };
  \end{feynman}
 \end{tikzpicture}\, \begin{tikzpicture}[baseline=(b3.base)]
 \begin{feynman}
    \vertex (a1);
    \vertex [below=1.0 of a1] (a2);
    \vertex [below=1.0 of a2] (a3);
    \vertex [below=1.0 of a3] (a4);
    \vertex [right=1.0 of a1] (b1);
    \vertex [below=1.0 of b1] (b2);
    \vertex [dot, below=1.0 of b2] (b3){};
    \vertex [below=1.0 of b3] (b4);
    \vertex [right=1.0 of b1] (c1);
    \vertex [below=1.0 of c1] (c2);
    \vertex [below=1.0 of c2] (c3);
    \vertex [below=1.0 of c3] (c4);

    \diagram* {
      (a1) -- [charged scalar, edge label'=\(p'\)] (b2),
      (b2) -- [photon, edge label=\(q'\)] (b3), 
      (b3) -- [anti charged scalar, edge label'=\(p\)] (a4),
      (c1) -- [anti charged scalar, edge label=\(k'\)] (b2),
      (b3) -- [charged scalar, edge label=\(k\)] (c4),
    };
  \end{feynman}
 \end{tikzpicture}\, \begin{tikzpicture}[baseline=(b3.base)]
 \begin{feynman}
    \vertex (a1);
    \vertex [below=1.0 of a1] (a2);
    \vertex [below=1.0 of a2] (a3);
    \vertex [below=1.0 of a3] (a4);
    \vertex [right=1.0 of a1] (b1);
    \vertex [dot, below=1.0 of b1] (b2){};
    \vertex [dot, below=1.0 of b2] (b3){};
    \vertex [below=1.0 of b3] (b4);
    \vertex [right=1.0 of b1] (c1);
    \vertex [below=1.0 of c1] (c2);
    \vertex [below=1.0 of c2] (c3);
    \vertex [below=1.0 of c3] (c4);

    \diagram* {
      (a1) -- [charged scalar, edge label'=\(p'\)] (b2),
      (b2) -- [photon, edge label=\(q'\)] (b3), 
      (b3) -- [anti charged scalar, edge label'=\(p\)] (a4),
      (c1) -- [anti charged scalar, edge label=\(k'\)] (b2),
      (b3) -- [charged scalar, edge label=\(k\)] (c4),
    };
  \end{feynman}
 \end{tikzpicture}& \nonumber\\
 \\
 &\begin{tikzpicture}[baseline=(b3.base)]
 \begin{feynman}
    \vertex (a1);
    \vertex [below=1.0 of a1] (a2);
    \vertex [below=1.0 of a2] (a3);
    \vertex [below=1.0 of a3] (a4);
    \vertex [right=1.0 of a1] (b1);
    \vertex [below=1.0 of b1] (b2);
    \vertex [below=1.0 of b2] (b3);
    \vertex [below=1.0 of b3] (b4);
    \vertex [right=1.0 of b1] (c1);
    \vertex [below=1.0 of c1] (c2);
    \vertex [below=1.0 of c2] (c3);
    \vertex [below=1.0 of c3] (c4);
    \vertex[right=1.0 of c1] (d1);
    \vertex[below=1.0 of d1] (d2);
    \vertex[below=1.0 of d2] (d3);
    \vertex[below=1.0 of d3] (d4);
    \vertex[below=0.25 of b2](b22);
    \vertex[right=0.25 of b22](e1);
    \vertex[above=0.25 of c3](e2);
    \vertex[left=0.25 of e2](e3);
    \vertex[below=0.25 of c2](e4);
    \vertex[left=0.25 of e4](e5);
    
    \diagram* {
      (a1) -- [charged scalar, edge label'=\(p'\)] (b2),
      (b2) -- [photon, edge label'=\(q''\)] (b3), 
      (b3) -- [anti charged scalar, edge label'=\(p\)] (a4),
      (b2) -- [scalar] (e1) -- [scalar, half left] (e3),
      (e3) -- [charged scalar, edge label=\(k\)] (d4),
      (d1) -- [anti charged scalar, edge label=\(k'\)] (e5),
      (b3) -- [scalar] (e5),
    };
  \end{feynman}  
 \end{tikzpicture}\, \begin{tikzpicture}[baseline=(b3.base)]
 \begin{feynman}
    \vertex (a1);
    \vertex [below=1.0 of a1] (a2);
    \vertex [below=1.0 of a2] (a3);
    \vertex [below=1.0 of a3] (a4);
    \vertex [right=1.0 of a1] (b1);
    \vertex [dot, below=1.0 of b1] (b2){};
    \vertex [below=1.0 of b2] (b3);
    \vertex [below=1.0 of b3] (b4);
    \vertex [right=1.0 of b1] (c1);
    \vertex [below=1.0 of c1] (c2);
    \vertex [below=1.0 of c2] (c3);
    \vertex [below=1.0 of c3] (c4);
    \vertex[right=1.0 of c1] (d1);
    \vertex[below=1.0 of d1] (d2);
    \vertex[below=1.0 of d2] (d3);
    \vertex[below=1.0 of d3] (d4);
    \vertex[below=0.25 of b2](b22);
    \vertex[right=0.25 of b22](e1);
    \vertex[above=0.25 of c3](e2);
    \vertex[left=0.25 of e2](e3);
    \vertex[below=0.25 of c2](e4);
    \vertex[left=0.25 of e4](e5);
    
    \diagram* {
      (a1) -- [charged scalar, edge label'=\(p'\)] (b2),
      (b2) -- [photon, edge label'=\(q''\)] (b3), 
      (b3) -- [anti charged scalar, edge label'=\(p\)] (a4),
      (b2) -- [scalar] (e1) -- [scalar, half left] (e3),
      (e3) -- [charged scalar, edge label=\(k\)] (d4),
      (d1) -- [anti charged scalar, edge label=\(k'\)] (e5),
      (b3) -- [scalar] (e5),
    };
  \end{feynman}  
 \end{tikzpicture}\, \begin{tikzpicture}[baseline=(b3.base)]
 \begin{feynman}
    \vertex (a1);
    \vertex [below=1.0 of a1] (a2);
    \vertex [below=1.0 of a2] (a3);
    \vertex [below=1.0 of a3] (a4);
    \vertex [right=1.0 of a1] (b1);
    \vertex [below=1.0 of b1] (b2);
    \vertex [dot, below=1.0 of b2] (b3){};
    \vertex [below=1.0 of b3] (b4);
    \vertex [right=1.0 of b1] (c1);
    \vertex [below=1.0 of c1] (c2);
    \vertex [below=1.0 of c2] (c3);
    \vertex [below=1.0 of c3] (c4);
    \vertex[right=1.0 of c1] (d1);
    \vertex[below=1.0 of d1] (d2);
    \vertex[below=1.0 of d2] (d3);
    \vertex[below=1.0 of d3] (d4);
    \vertex[below=0.25 of b2](b22);
    \vertex[right=0.25 of b22](e1);
    \vertex[above=0.25 of c3](e2);
    \vertex[left=0.25 of e2](e3);
    \vertex[below=0.25 of c2](e4);
    \vertex[left=0.25 of e4](e5);
    
    \diagram* {
      (a1) -- [charged scalar, edge label'=\(p'\)] (b2),
      (b2) -- [photon, edge label'=\(q''\)] (b3), 
      (b3) -- [anti charged scalar, edge label'=\(p\)] (a4),
      (b2) -- [scalar] (e1) -- [scalar, half left] (e3),
      (e3) -- [charged scalar, edge label=\(k\)] (d4),
      (d1) -- [anti charged scalar, edge label=\(k'\)] (e5),
      (b3) -- [scalar] (e5),
    };
  \end{feynman}  
 \end{tikzpicture}\,  \begin{tikzpicture}[baseline=(b3.base)]
 \begin{feynman}
    \vertex (a1);
    \vertex [below=1.0 of a1] (a2);
    \vertex [below=1.0 of a2] (a3);
    \vertex [below=1.0 of a3] (a4);
    \vertex [right=1.0 of a1] (b1);
    \vertex [dot, below=1.0 of b1] (b2){};
    \vertex [dot, below=1.0 of b2] (b3){};
    \vertex [below=1.0 of b3] (b4);
    \vertex [right=1.0 of b1] (c1);
    \vertex [below=1.0 of c1] (c2);
    \vertex [below=1.0 of c2] (c3);
    \vertex [below=1.0 of c3] (c4);
    \vertex[right=1.0 of c1] (d1);
    \vertex[below=1.0 of d1] (d2);
    \vertex[below=1.0 of d2] (d3);
    \vertex[below=1.0 of d3] (d4);
    \vertex[below=0.25 of b2](b22);
    \vertex[right=0.25 of b22](e1);
    \vertex[above=0.25 of c3](e2);
    \vertex[left=0.25 of e2](e3);
    \vertex[below=0.25 of c2](e4);
    \vertex[left=0.25 of e4](e5);
    
    \diagram* {
      (a1) -- [charged scalar, edge label'=\(p'\)] (b2),
      (b2) -- [photon, edge label'=\(q''\)] (b3), 
      (b3) -- [anti charged scalar, edge label'=\(p\)] (a4),
      (b2) -- [scalar] (e1) -- [scalar, half left] (e3),
      (e3) -- [charged scalar, edge label=\(k\)] (d4),
      (d1) -- [anti charged scalar, edge label=\(k'\)] (e5),
      (b3) -- [scalar] (e5),
    };
  \end{feynman}  
 \end{tikzpicture}&
 \nonumber
\end{eqnarray} Here, the four upper diagrams correspond to the t-channel of the interaction, while the four lower diagrams correspond to the u-channel. We must further emphasize that the particle-flow arrows illustrated in Eq. \eqref{graphmmamma111222} are distinct from those depicted in Eqs. \eqref{graffeyntreelschannel} and \eqref{graphmamammama12345}. This distinction arises in this section because we are dealing solely with particles (and not antiparticles). Having said that, by standard procedure, we can write the thermal scattering matrix for this case as

\begin{eqnarray}\label{mhatbetamubetamenosmtbeta}
    \hat{\mathcal{M}}(\beta) = \hat{\mathcal{M}}_u(\beta) - \hat{\mathcal{M}}_t(\beta),
\end{eqnarray} with
\begin{eqnarray}
    \hat{\mathcal{M}}_u(\beta) &=& \frac{e^2}{q''^{2}} \left(\prod v_B\right)\left(1+\prod \frac{u_B}{v_B}\right)\left\lbrace 1+2 \pi i\,  v_B^2(\mathbf{q''},\beta)\, q''^2\, \delta(q''^2)\left[ \frac{1-\prod \frac{u_B}{v_B}}{1+\prod \frac{u_B}{v_B}} \right]\right\rbrace  \nonumber\\
    &\times& \left\lbrace \left( k'\cdot p' + k\cdot k'+k\cdot p+p\cdot p'\right) + 2\frac{g}{e}\, \epsilon^{\mu \nu \alpha \beta}\, \omega_{\nu}\, q''_{\alpha}\, (k_{\beta}+p'_{\beta})(k'_{\mu}+p_{\mu}) \right.\nonumber\\
    &-& \left. \frac{g^2}{e^2}\, \epsilon^{\mu \nu \alpha \beta}\, \epsilon^{\gamma \rho \eta \sigma}\, \eta_{\beta \sigma}\, \omega_{\nu}\, \omega_{\rho}\, q''_{\alpha}\, q''_{\eta}\, (k_{\gamma}+p'_{\gamma})(k'_{\mu}+p_{\mu})\right\rbrace,
\end{eqnarray} and $\hat{\mathcal{M}}_t(\beta)$ still given by Eq. \eqref{matrixchannelthatbeta}. As in the previous sections, once Eq. \eqref{mhatbetamubetamenosmtbeta} is known, we must proceed by evaluating an expression for the thermal differential cross section in the two cases of interest: when $\omega^{\mu}$ is a timelike or spacelike four-vector.

\subsubsection{Timelike case}

\begin{figure}[!]
     \centering
     \begin{subfigure}[b]{0.3\textwidth}
         \centering
         \includegraphics[scale=0.6]{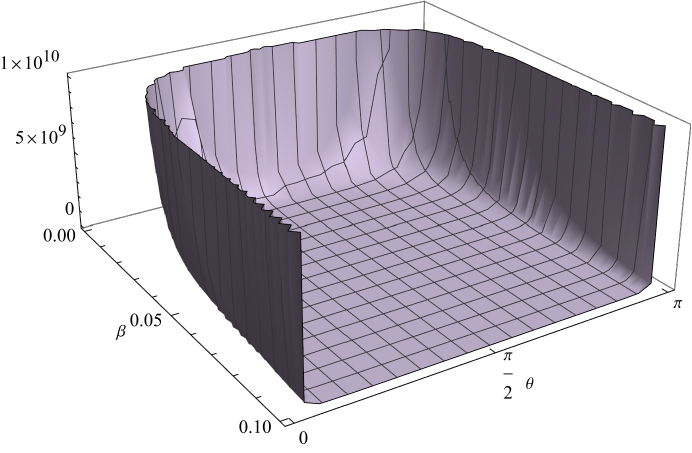}
         \caption{}
         \label{difsectime3a}
     \end{subfigure}
     \hfill
     \begin{subfigure}[b]{0.6\textwidth}
         \centering
         \includegraphics[scale=0.6]{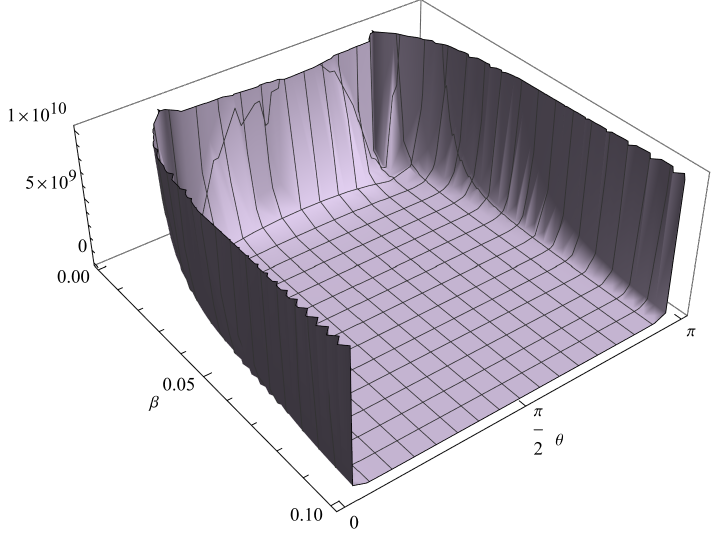}
         \caption{}
         \label{difsectime3b}
     \end{subfigure}
        \caption{\raggedright Angular and temperature dependencies of the differential cross section for $a$-type meson-meson scattering in $a$-type meson-meson in the timelike case. For this plot we took $e=1$ and  $E_{cm} = 1$. In \eqref{difsectime3a}, no Lorentz violation has been considered ($g \omega_0=0$). In \eqref{difsectime3b}, we have set $g \omega_0=5$.}
        \label{fig6timelikediseccccc}
\end{figure}

As before, $\omega^{\nu} = (\omega_0,\mathbf{0})$ for this case. Thus, the temperature-dependent differential cross section for mesons of $a$-type scattering off mesons of $a$-type takes the form

\begin{eqnarray}\label{difsecthermaltimelikeccc}
 \left( \frac{d\sigma}{d\Omega} \right)_{\beta,cm} &=& \frac{e^4}{256\, \pi ^2\, E_{cm}^2} \Bigg\{ (\cos (2 \theta )+7)^2 \csc ^4(\theta ) \, C(\beta)\nonumber\\
 &+& \pi^2\, D(\beta)\, \big[ (\cos (\theta )+3)^2 \delta \left(E_{cm}^2 (\cos (\theta )-1)\right)^2\nonumber\\
&-& (\cos (2 \theta )-17) \delta \left(E_{cm}^2 (\cos (\theta )+1)\right) \delta \left(E_{cm}^2 (\cos (\theta
   )-1)\right)\nonumber\\
   &+& (\cos (\theta )-3)^2 \delta \left(E_{cm}^2 (\cos (\theta )+1)\right)^2 \big] \Bigg\}\nonumber\\
   &-& \frac{1}{128} \left( \frac{e\, g\, \omega_0}{\pi} \right)^2 \Bigg\{ (\cos (2 \theta )+7) \csc ^2(\theta )\, C(\beta)\nonumber\\
   &+& \frac{\pi^2}{2}\, D(\beta)\, \sin^2 (\theta) \Big[(\cos (\theta )+3) \delta \left(E_{cm}^2 (\cos (\theta )-1)\right)^2\nonumber\\
   &-&(\cos (\theta )-3) \delta \left(E_{cm}^2 (\cos (\theta
   )+1)\right)^2+6 \delta \left(E_{cm}^2 (\cos (\theta )-1)\right) \delta \left(E_{cm}^2 (\cos (\theta )+1)\right) \Big] \Bigg\}\nonumber\\
   &+& \frac{\pi^2\, E_{cm}^2}{256}\, \left( \frac{g\, \omega_0}{\pi} \right)^4 \Bigg\{ C(\beta) + \frac{\pi^2}{4}\, D(\beta)\, \sin^4(\theta)\, \Big[ \delta \left(E_{cm}^2 (\cos (\theta )-1)\right)^2\nonumber\\
   &+& \delta \left(E_{cm}^2 (\cos
   (\theta )+1)\right)^2 + 2\, \delta \left(E_{cm}^2 (\cos (\theta )-1)\right) \delta \left(E_{cm}^2 (\cos (\theta
   )+1)\right) \Big] \Bigg\}.
\end{eqnarray} Once again, we observe that first and third-order terms in the coupling constant $g$ are not present when Lorentz symmetry violation is characterized by a spacetime quadrivector $\omega^{\mu}=(\omega_0,\mathbf{0})$, completely corroborating what was previously discussed in Sec. \ref{secmamatypemamatype}. The temperature dependence is, as before, governed by functions $C(\beta)$ and $D(\beta)$. Figure \eqref{fig6timelikediseccccc} illustrates the angular and temperature dependencies of the differential cross section found in Eq. \eqref{difsecthermaltimelikeccc} for the cases with and without Lorentz violation. Once again, it is observed that the effects of anisotropy appear to be more pronounced in the limit of very high temperatures. This corroborates findings in Sec. \ref{secmamatypemamatype} and further emphasizes the need for experimental investigations in this regime of temperature.

\subsubsection{Spacelike case}

\begin{figure}[!]
     \centering
     \begin{subfigure}[b]{0.3\textwidth}
         \centering
         \includegraphics[scale=0.6]{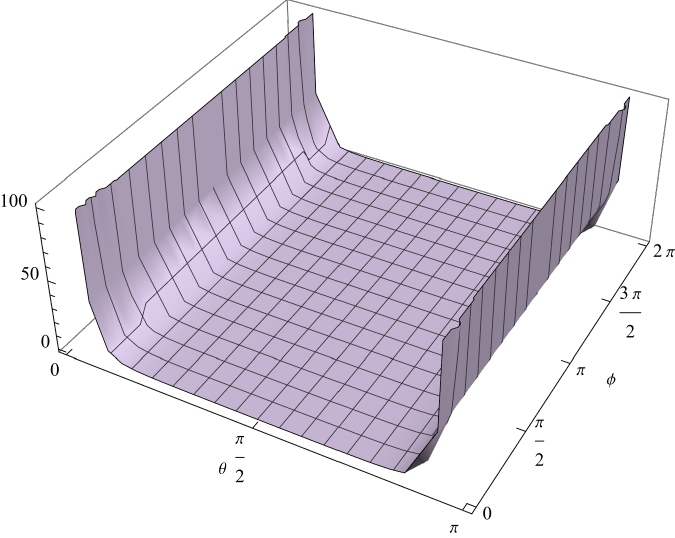}
         \caption{}
         \label{difsecspace3a}
     \end{subfigure}
     \hfill
     \begin{subfigure}[b]{0.6\textwidth}
         \centering
         \includegraphics[scale=0.6]{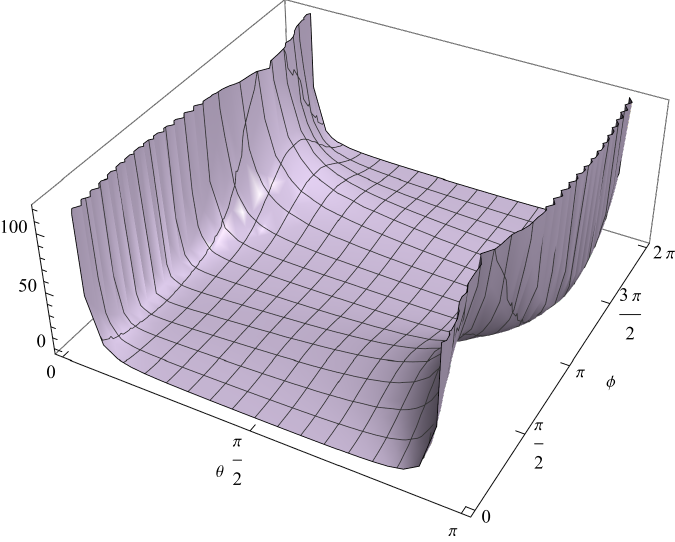}
         \caption{}
         \label{difsecspace3b}
     \end{subfigure}
     \hfill
     \begin{subfigure}[b]{0.6\textwidth}
         \centering
         \includegraphics[scale=0.6]{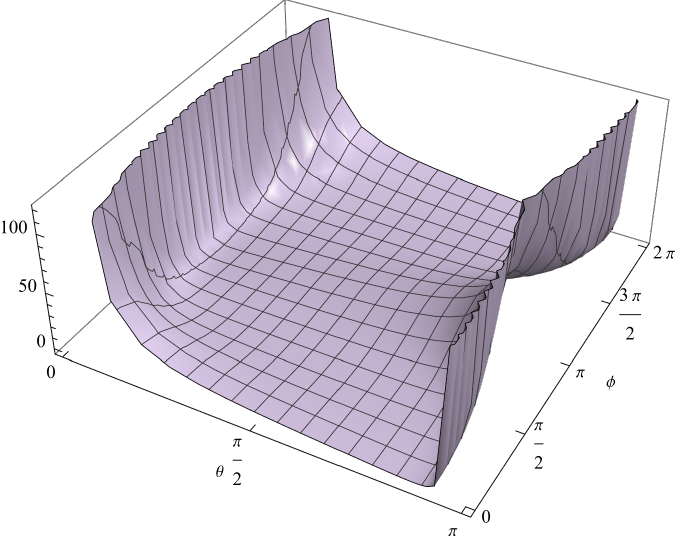}
         \caption{}
         \label{difsecspace3c}
     \end{subfigure}
        \caption{\raggedright Angular dependencies of the zero-temperature differential cross section for $a$-type meson-meson scattering in $a$-type meson-meson in the spacelike case. For this plot we took $e=1$,  $E_{cm} = 1$. In \eqref{difsecspace3a}, we have the usual case ($g|\mathbf{w}| =0$). In \eqref{difsecspace3b} and \eqref{difsecspace3c}, we have considered $g|\mathbf{w}| =5$, but for the former $\theta_{\omega z}=\frac{\pi}{2}$ and $\phi_{\omega x}=0$, while for the latter $\theta_{\omega z}=\frac{\pi}{2}$ and $\phi_{\omega x}=\frac{\pi}{2}$.}
        \label{fig7difsecspacecccccultcross}
\end{figure}


As we know, the violation of Lorentz symmetry is characterized by the spacetime quadrivector $\omega^{\mu}=(0,\mathbf{w})$. Thus, for this case, the thermal differential cross section can be written as 
\begin{eqnarray}
   \left( \frac{d\sigma}{d\Omega} \right)_{\beta,cm} &=&  \frac{e^4 }{256 \pi ^2 E_{cm}^2} \Bigg\{ (\cos (2 \theta )+7)^2 \csc ^4(\theta ) \, C(\beta)\nonumber\\
   &+& \pi^2 D(\beta) \Big[ (\cos (\theta )+3)^2 \delta \left(E_{cm}^2 (\cos (\theta )-1)\right)^2 + (\cos (\theta )-3)^2 \delta \left(E_{cm}^2
   (\cos (\theta )+1)\right)^2\nonumber\\
   &-& (\cos (2 \theta )-17) \delta \left(E_{cm}^2 (\cos (\theta )-1)\right) \delta \left(E_{cm}^2 (\cos (\theta
   )+1)\right) \Big] \Bigg\}\nonumber\\
   &-& g\, |\mathbf{w}|\, \frac{e^3}{32\, \pi ^2\, E_{cm}}\Bigg\{ (15 \cos (\theta )+\cos (3 \theta )) \csc ^3(\theta ) \sin \left(\theta _{\omega z}\right) \sin
   \left(\phi -\phi _{\omega x}\right)\, C(\beta)\nonumber\\
   &+& \frac{\pi^2}{2}\, D(\beta) \Big[ (6 \sin (\theta )+\sin (2 \theta )) \sin \left(\theta _{\text{$\omega
   $z}}\right) \sin \left(\phi -\phi _{\omega x}\right) \delta \left(E_{cm}^2 (\cos (\theta )-1)\right)^2 \nonumber\\
   &+& (\sin (2 \theta )-6 \sin (\theta )) \sin \left(\theta _{\omega z}\right) \sin \left(\phi -\phi _{\text{$\omega
   $x}}\right) \delta \left(E_{cm}^2 (\cos (\theta )+1)\right)^2 \nonumber\\
   &-& 2 \sin (2 \theta ) \sin \left(\theta _{\omega z}\right) \sin \left(\phi -\phi _{\omega x}\right) \delta \left(E_{cm}^2 (\cos (\theta )-1)\right) \delta \left(E_{cm}^2 (\cos
   (\theta )+1)\right) \Big] \Bigg\} \nonumber\\
   &+& \cdots ,
\end{eqnarray} where the dots stand for higher-order Lorentz-breaking terms, and once more, since the exact form of these corrections is lengthy, they will not be displayed in detail here. However, it is important to emphasize that all the terms up to fourth order in $g$ are once again present, further corroborating the suggestion that the scattering processes involving mesons and antimesons of the same type are more sensitive to Lorentz violation effects characterized by a spacelike four-vector within the context of the QED framework here discussed. The thermal behavior is once again shaped by the functions $C(\beta)$ and $D(\beta)$, resulting in a significant differential cross section value at very high temperature limits. The zero-temperature result is also promptly recovered, given that $C(\beta)\rightarrow 4$ and $D(\beta) \rightarrow 0$ (see Figure \eqref{canddgraph}). The angular dependencies of the zero-temperature differential cross section for some values of $\theta_{\omega z}$ and $\phi_{\omega x}$ are illustrated in Figure \eqref{fig7difsecspacecccccultcross}, revealing that the anisotropy effects are most pronounced at the boundaries of the interval to which $\theta$ belongs.


%

\section{Conclusion}\label{conclusion}
 
In this paper we study meson scattering in a non-minimally Lorentz-violating scalar QED at finite temperature. The meson scatterings were investigated in tree level and the finite temperature effects were addressed by using the thermofield dynamics formalism. We have considered three types of scattering, namely, meson-antimeson of $a$-type into meson-antimeson of $b$-type, meson-antimeson of $a$-type into meson-antimeson of $a$-type and meson-meson of $a$-type into meson-meson of $a$-type. All cases considered were addressed similarly and for each scattering we have computed the cross section in order to investigate the influence of the Lorentz symmetry violation and finite temperature effects.

For the meson-antimeson of $a$-type into meson-antimeson of $b$-type, we have considered contributions up to second order in the Lorentz-violating parameter. Also, we have separately investigated two distinct possibilities based on whether $\omega^{\mu}$ is a timelike or spacelike four-vector. For the timelike case, no Lorentz violation effect in the cross section is observed but the temperature influence becomes highly relevant in the limit of high temperatures. For the spacelike case, unlike the timelike case, Lorentz violation effects are observed, suggesting that the scattering in question is more sensitive in scenarios where $\omega^{\mu}$ is a spacelike four-vector. Also, such effects are completely canceled for situations where $\mathbf{w}$ lies in the plane containing the line defining the collision trajectory of the particles at the beginning of the process. 

For the meson-antimeson of a-type into meson-antimeson of a-type, the timelike consideration shows us that the scattering process is more sensitive to Lorentz violation effects since these are dependent of the parameter $\omega_0$, i.e., the timelike four-vector $\omega^{\mu}$. It is important to highlight that we observed that neither first-order nor third-order terms in the coupling constant $g$ are present in the cross section. This seems to be a characteristic of scattering involving mesons and antimesons of the same species within the context of the non-minimally Lorentz-violating scalar QED framework presented in this work. Also, the effects of anisotropy characterized by the parameter $\omega_0$ become more pronounced in the limit of very high temperatures, suggesting a new perspective (or path) in the search for Lorentz-violating measurable effects. A similar behaviour occurs for the spacelike consideration, but in this case all the terms up to fourth order in $g$ are present.

Finally, for the meson-meson of $a$-type into meson-meson of $a$-type, we have similar corrections due to the Lorentz symmetry violation and temperature in comparison with the the meson-antimeson of a-type into meson-antimeson of a-type process, being the coefficients of each order of contribution the sole difference between the two cases. An interesting behaviour present in all three processes is the angular dependency promoted by the Lorentz symmetry violation regarding the vector $\mathbf{w}$ that sets the priviledge space direction and the line defining the collision trajectory of the particles at the beginning of the process. Such angular dependency can be seen as an effective way of verifing the existence of a priviledge direction in spacetime, by the measurement of two different cross sections for the same process occuring in different directions.


\section*{Acknowledgments}
\hspace{0.5cm} JF would like to thank the Funda\c{c}\~{a}o Cearense de Apoio ao Desenvolvimento Cient\'{i}fico e Tecnol\'{o}gico (FUNCAP) under the grant PRONEM PNE0112-00085.01.00/16 for financial support, the National Council for scientific and technological support (CNPq) under the grant 304485/2023-3 and Gazi University for the kind hospitality. RVM would like to thank  CNPq under the Grant no. 200879/2022-7 for financial support. 
\vspace{0.5cm}
\center{\bf No Data associated in the manuscript}

\appendix


\end{document}